\documentclass[12pt]{article}
\usepackage{epsf}
\global\arraycolsep=2pt 

\begin{document}
\makeatletter
\def\fmslash{\@ifnextchar[{\fmsl@sh}{\fmsl@sh[0mu]}}
\def\fmsl@sh[#1]#2{%
  \mathchoice
    {\@fmsl@sh\displaystyle{#1}{#2}}%
    {\@fmsl@sh\textstyle{#1}{#2}}%
    {\@fmsl@sh\scriptstyle{#1}{#2}}%
    {\@fmsl@sh\scriptscriptstyle{#1}{#2}}}
\def\@fmsl@sh#1#2#3{\m@th\ooalign{$\hfil#1\mkern#2/\hfil$\crcr$#1#3$}}
\makeatother
\newcommand{\into}[1]{\mbox{d}{#1}\enspace}
\newcommand{\intf}[1]{\mbox{d}^4{#1}\enspace}
\newcommand{\be}{\begin{eqnarray}}
\newcommand{\ee}{\end{eqnarray}}
\newcommand{\ben}{\begin{equation}}
\newcommand{\een}{\end{equation}}
\newcommand{\tr}[1]{\mbox{Tr}\left\{#1\right\}}
\newcommand{\trans}{^{\mbox{{\scriptsize T}}}}
\newcommand{\abl}[2]{\frac{\mbox{d}#1}{\mbox{d}#2}}
\newcommand{\pabl}[2]{\frac{\partial #1}{\partial #2}}
\newcommand{\unter}[1]{_{\mbox{\scriptsize #1}}}
\newcommand{\untter}[1]{_{\mbox{\tiny #1}}}
\newcommand{\ex}[1]{\mbox{e}^{#1}}
\newcommand{\im}[1]{\mbox{ Im}{#1}}
\newcommand{\Det}[1]{\mbox{det}\left({#1}\right)}
%
\thispagestyle{empty}
\begin{titlepage}

\begin{flushright}
TTP 98--20 \\
\today
\end{flushright}
\vspace{0.3cm}
\boldmath
\begin{center}
\Large\bf Improved Unitarity Bounds for $\bar B^0 \to \pi^+ \ell^-
\bar\nu_\ell$ Decays
\end{center}
\unboldmath
\vspace{0.8cm}

\begin{center}
{\large Thomas Mannel} and {\large Boris Postler} \\
{\sl Institut f\"{u}r Theoretische Teilchenphysik,Universit\"at Karlsruhe,}\\
{\sl     D -- 76128 Karlsruhe, Germany.} \\
\end{center}

\vspace{\fill}

\begin{abstract}
\noindent
We obtain model independent bounds for the form factors which arise in
semileptonic $B\to\pi$ decays. To this end
we derive a theoretical restriction for possible combinations of the value 
of the form factor and its derivatives at the end point region. These
restrictions are then used as an input in a --- slightly modified --- general 
formalism, which has already been deduced and applied in earlier publications.
Our results can be used to set constraints on the form factors which have to
be obeyed by all model dependent parametrizations.
\end{abstract}
\end{titlepage}

\section{Introduction}
One important parameter for the determination of the unitarity triangle
of the CKM matrix is $V_{ub}$. While its phase can only be determined 
from non--leptonic decays, its magnitude will be determined from 
semileptonic processes. Various methods have been proposed to extract 
$V_{ub}$ from both inclusive as well as exclusive semileptonic decays, 
and the final numbers after the era of the $B$ factories will probably 
be obtained from a mixture of both exclusive and inclusive methods. 

The disadvantage of exclusive decays is their dependence on form factors
which are nonperturbative quantities. Modelling these form factors 
necessarily introduces some uncontrollable systematic uncertainty into 
the determination of $V_{ub}$ and hence a model independent method is
desirable. At present the values of $V_{ub}$ extracted from exclusive 
semileptonic $B\to\pi$ decays using lattice QCD, QCD sum rules and 
quark models range from $|V_{ub}| = (2.5 - 4.5)\times10^{-3}$. Clearly
this is not a satisfactory situation, in particular in view of the 
data to be expected from the $B$ factories.  Although there have been
recent efforts to update existing models (see
e.g. \cite{hphx9711268}--\nocite{hphx9712399,hphx9801421}\cite{hphx9801443}), 
there is still room for some improvement.

There are constraints on form factors originating from their analyticity 
properties and the unitarity of the underlying theory. The ideas 
how to implement these constraints are in fact quite old 
\cite{prvxd4x3519}--\nocite{prvxd3x2807,prvxd4x725,prvxd4x2020}\cite{npxb189x157}
and have recently attracted renewed attention. In particular, combining these
unitarity bounds with lattice data \cite{hphx9509358} has lead to relatively
tight and model independent bounds on the form factors of $B \to \pi$
transitions.

In the present paper we improve the application of unitarity bounds 
in such a way that not only points at which the form factor is known
(e.g. from lattice data or heavy quark relations) can be included, 
but also the slope and higher derivatives at some point, 
which could be known for instance from 
sum rule considerations. At the point of maximal 
momentum transfer the derivatives are correlated with the value of the 
form factor due to analyticity and we shall discuss the restrictions
obtained from this in some detail.   

A similar improvement of the unitarity bounds has been proposed by
Boyd et al.~\cite{hphx9702300,hphx9705252}, where the form factors for 
$B \to \pi \ell \nu$ are expanded in a cleverly chosen conformal 
variable $z$ and some of the relations we use in the present paper appear 
in a similar form in these references. 

In the next section we shall summarize the formalism as described in 
\cite{npxb189x157}.
The improved bounds are considered in section \ref{secimprov} where we
generalize the method to include known slopes and higher 
derivatives. We make use of the correlation between the derivatives 
and the value of the form factor taken at $q^2_{max}$ to tighten 
the bounds even without using any additional physics input. In order 
to get useful bounds some physics input is necessary and we chose to 
use the chiral limit which is valid close to the end point. This has 
a strong effect since any knowledge of the form factors in the 
end point region tightens the bounds significantly.  
With this input we study numerical examples. 
\section{Bounds on Form Factors}
\label{secbounds}
For later use we summarize in this section the formalism how to derive 
bounds on the form factors using only perturbative QCD, unitarity and 
the ana\-ly\-ti\-ci\-ty of the form factors in the complex plane. 

We choose to describe the hadronic matrix element of the semileptonic 
$\bar B^0 \to \pi^+ \ell^- \bar\nu_\ell$ decays with the following two 
form factors:
\be
\label{anfang}
\langle \pi^+(p')|V^\mu|\bar B^0(p)\rangle
&=&\left(p^\mu+{p'}^\mu-\frac{M^2-m^2}{q^2}q^\mu\right)f^+(q^2)\\
&&+\frac{M^2-m^2}{q^2}q^\mu f^0(q^2)\nonumber
\ee
where $M^2=m_B^2$, $m^2=m_\pi^2$, $V^\mu=\bar u\gamma^\mu b$ and $q=p-p'$.
In this notation $q^2$ runs from $q^2_{min}=m_\ell^2$ to
$q^2_{max}=(M-m)^2$. Throughout this paper we will neglect the lepton masses
and therefore set $q^2_{min}=0$. Furthermore, this special choice of
decomposition into Lorentz vectors leads to form factors which have to satisfy
the kinematical constraint
\ben
\label{kinconst}
f^+(0) = f^0(0).
\een
The form factors are real functions of a real variable but it is
convenient to think of them as analytic functions in the complex
$q^2$--plane. 

To derive bounds on $f^+(q^2)$ and $f^0(q^2)$ we consider the
two--point function 
\be
\label{2pfunction}
\Pi^{\mu\nu} &\equiv& i\int\intf{x}\ex{iq\cdot
x}\langle0|T\{V^\mu(x)V^{\nu\dagger}(0)\}|0\rangle \nonumber\\
&\equiv& -(g^{\mu\nu}q^2 - q^\mu q^\nu)\Pi_T(q^2) + q^\mu q^\nu \Pi_L(q^2),
\ee
where $V^\mu=\bar u\gamma^\mu b$ as above and with
$\Pi_{T/L}(q^2)$ corresponding to the propagation of a $J^P=1^-/0^+$
particle. This two--point function can be (and has been) reliable evaluated
in the deep euclidean region where $-q^2 \equiv Q^2 \gg \Lambda^2\unter{QCD}$,
i.e. at an energy scale where perturbative QCD is appliccable.

Including a sum over all possible intermediate states
$\Gamma$ we get the following result for the imaginary part of
 (\ref{2pfunction}): 
\be
\label{states}
-(g^{\mu\nu} q^2 - q^\mu q^\nu)\im{\Pi_T(q^2)}+q^\mu q^\nu\im{\Pi_L(q^2)}
\nonumber \\
= \frac 12\sum_\Gamma(2\pi)^4 \delta^{(4)}(q-p_\gamma)\langle0|V^\mu(0)|\Gamma
\rangle\langle\Gamma|V^{\nu\dagger}(0)|0\rangle.
\ee
We therefore get an equation for the spectral functions $\im{\Pi_{T/L}(q^2)}$
(i.e. for the absorptive parts of $\Pi_{T/L}(q^2)$) which we can relate to the
real parts using the substracted dispersion relations
\be
\label{displ}
\chi_L(Q^2)\equiv\left(-\pabl{}{Q^2}\right)(-Q^2 \Pi_L(Q^2)) =\frac
1\pi\int_0^\infty\into{t} \frac{t\im{\Pi_L(t)}}{(t+Q^2)^2}
\ee
and
\be
\label{dispt}
\chi_T(Q^2)\equiv\frac12\left(-\pabl{}{Q^2}\right)^2(-Q^2 \Pi_T(Q^2)) =\frac
1\pi\int_0^\infty\into{t} \frac{t\im{\Pi_T(t)}}{(t+Q^2)^3}.
\ee

We now restrict ourselves to include only contributions of the $B^*$ and
the $B\pi$ states in the sum over all intermediate states in
 (\ref{states}).
It is possible to discard all other intermediate states because 
 (\ref{states}) is a sum of positive terms if the indices are treated
symmetrically. Using isospin and crossing symmetry we can use
(\ref{anfang}) to express (\ref{states}) by the two form factors.
Projecting out now the transversal/longitudinal parts, one gets the
inequalities
\be
\label{ineql}
\im{\Pi_L(t)}\ge\frac 32\frac{t_+ t_-}{16\pi}\sqrt{(t-t_+)(t-t_-)}\frac
{|f^0(t)|^2}{t^3}\Theta(t-t_+)
\ee
and
\be
\label{ineqt}
\im{\Pi_T(t)} &\ge&\pi\left(\frac{m_{B^*}}{f_{B^*}}\right)^2
\delta(t-m_{B^*}^2) \nonumber\\
&+& \frac32 \frac
1{48\pi}\frac{[(t-t_+)(t-t_-)]^{3/2}}{t^3}|f^+(t)|^2\Theta(t-t_+),
\ee
where $t=q^2$ and $t_\pm = (M\pm m)^2$.
 
It is now possible to get bounds on the form factors if one inserts
(\ref{ineql},\ref{ineqt}) in (\ref{displ},\ref{dispt}). Since the l.h.s. of 
(\ref{displ},\ref{dispt}) can be calculated for $Q^2 \gg \Lambda^2\unter{QCD}$
using perturbative QCD one gets inequalities which restrict the form factors. 
These inequalities take the form (in shorthand notation) 
\be
\label{shorthand}
J(Q^2) \ge \frac 1\pi \int_{t_+}^\infty\into{t} k(t,Q^2)|f(t)|^2
\ee
where $J(Q^2)$ denotes the QCD input (i.e. the perturbative calculation of
$\chi_{T/L}$) including the $B^*$--resonance in case of $f^+$
and $k(t,Q^2)$ is a known kinematical function. The exact value/structure of 
$J(Q^2)$ and $k(t,Q^2)$ does of course depend on the form factor under
examination.

To translate the inequality (\ref{shorthand}) in constraints on the form factor
for values of $t$ in the range $[0,t_-]$ (which is the kinematical region of 
physical interest), we map the complex $t$--plane into the unit disc with 
the conformal transformation
\be
\frac{1+z}{1-z} = \sqrt{\frac{t_+-t}{t_+-t_-}}
\ee
so that (\ref{shorthand}) becomes
\be
\label{main}
J(Q^2)\ge\int_{|z|=1} \frac{\mbox{d}z}{2\pi i z}|\phi(z,Q^2) f(z)|^2
\ee
with $f(z)\hat = f(t(z))$.
Here we have used the fact, that $k(t,Q^2)$ is a positive definite
quantity so that $\phi(z,Q^2)\equiv\sqrt{k(t(z),Q^2)}$ times the squareroot
of the Jacobian of the transformation. 

The value of the form factor $f(z)$ for any point $z(t)$ is accessible 
by defining an inner product
\be
\langle g|h \rangle = \int_{|z|=1} \frac{\mbox{d}z}{2\pi i z} \bar g(z)h(z)
\ee
and by considering the product $\langle g_t|\phi f\rangle$ where
\be
g_t = \frac 1{1-\bar z(t)z},
\ee
so that $f(z(t))$ has no poles in the range $[0,t_-]$. Cauchy's 
theorem now yields
\be
\label{einf}
\langle g_t|\phi f\rangle = \phi(z(t),Q^2) f(z(t)).
\ee
On the other hand, if there is a pole at $t=t_p$ away from the cut in the
complex $t$--plane (i.e. for $t>t_-$), one would obtain
\be
\langle g_t|\phi f\rangle = \phi(z(t),Q^2) f(z(t)) + \frac{\mbox{Res}
\{\phi f,z(t_p)\}}{z(t_p)-z(t)}.
\ee
The residue can either be approximated or eliminated completely by the 
transformation \cite{plxb301x257}
\be
\label{phitrafo}
\phi(z,Q^2)\quad \to \quad \phi_p(z,Q^2)\equiv \phi(z,Q^2)\frac {z-z(t_p)}
{1-\bar z(t_p)z}
\ee
where $t_p$ is assumed to lie in the range $[t_-,t_+]$ so that $\phi_p$ is 
positive for $z=z(t)$ with $t$ in $[0,t_-]$. This replacement cancels the
pole of $f(t)$ so that (\ref{einf}) holds. The crucial property of this
transformation is the fact that, since $|(z-z(t_p))/(1-\bar z(t_p)z)| =1$
for $z$ on the unit circle, $\langle \phi f|\phi f\rangle=\langle \phi_p
f|\phi_p f\rangle$ and the QCD constraints are left unchanged.

\begin{figure}[ht]
\epsfxsize=12cm
\leavevmode
\centering
\epsffile[70 250 540 540]{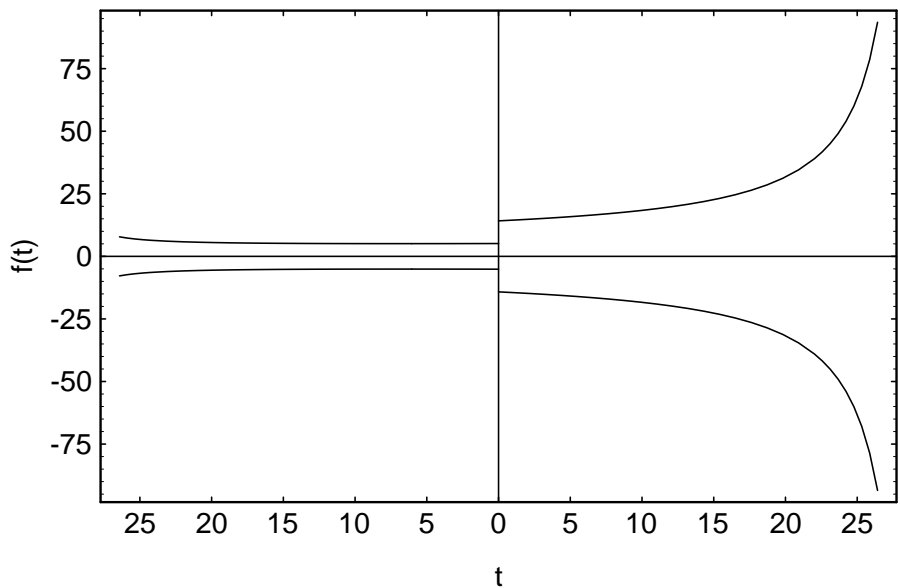} \\
\caption{Bounds on the form factors $f^0(t)$ (on the left side, $t$ increasing 
to the left) and $f^+(t)$ (on the right side), derived
without any additional constraints and plotted over the whole kinematical
range of $[0,t_-]$ ($t$ in $\mbox{GeV}^2$).
\label{figwithout}} 
\end{figure}

Because of the positivity of the inner product we have
\be
\Det{
\begin{array}{cc}
\langle \phi f |\phi f \rangle & \langle \phi f |g_t \rangle \\
\langle g_t |\phi f \rangle & \langle g_t |g_t \rangle
\end{array}
} \ge 0,
\ee
which, by eliminating $\langle\phi f|\phi f\rangle$ with (\ref{main}), 
gives us the following bounds on the form factors
\be
\label{withoutpoints}
|f(t)|^2 \le J(Q^2)\frac1{1-z^2(t)}\frac1{|\phi(z(t),Q^2)|^2}
\ee
where we have made use of the fact that $z(t)$ is real for $t$ in $[0,t_-]$.

These Bounds, which are derived using only perturbative QCD and which do not
depend on any model assumptions or input parameters (they are therefore quite
loose), are plotted in figure \ref{figwithout}. They were calculated using
$Q^2=0$ in $J(Q^2)$ at order ${\cal O}(\alpha_s)$ 

If we know the value of the form factor $f_{i}\equiv f(t_{i})$
at $n$ points $i=1,\dots, n$, we can define a matrix $\cal{M}$
\be
\label{mat}
\cal{M} = \left(
\begin{array}{ccccc}
\langle \phi f |\phi f \rangle & 
\langle \phi f |g_t \rangle &
\langle \phi f |g_{t_1} \rangle &
\cdots &
\langle \phi f |g_{t_n} \rangle \\

\langle g_t |\phi f \rangle & 
\langle g_t |g_t \rangle &
\langle g_t |g_{t_1} \rangle &
\cdots &
\langle g_t |g_{t_n} \rangle \\

\langle g_{t_1} |\phi f \rangle & 
\langle g_{t_1} |g_t \rangle &
\langle g_{t_1} |g_{t_1} \rangle &
\cdots &
\langle g_{t_1} |g_{t_n} \rangle \\

\vdots & \vdots & \vdots & \ddots & \vdots \\

\langle g_{t_n} |\phi f \rangle & 
\langle g_{t_n} |g_t \rangle &
\langle g_{t_n} |g_{t_1} \rangle &
\cdots &
\langle g_{t_n} |g_{t_n} \rangle
\end{array}
\right)
\ee
and obtain, using again the positivity of the inner product,
\be
\label{ineq}
\det{\cal M} \ge 0.
\ee
Since $\langle \phi f |\phi f \rangle$ is eliminated using (\ref{main})
and since all components except $f(t)$ 
of this matrix are known --- they are either constants or functions of $t$ ---
the inequality (\ref{ineq}) leads to bounds on the
form factors $f(t)$. The inclusion of some known values  (which i.e. are 
predicted by a model or which one can get by lattice calculations, see
\cite{hphx9509358}) therefore will give us more stringent bounds 
\be
F_{lo}(t,t_i, f_i) \le f(t) \le F_{up}(t,t_i, f_i)
\ee
with $F_{lo}$ and $F_{up}$ calculable functions of $t$ with the parameters 
$t_i,f_i$.

The upper and lower bounds can be written as
\be
\label{bounds}
F_{up,lo}(t,Q^2) =
\frac{-\beta(t)\pm\sqrt{c(Q^2)\cdot\Delta(t)}}{\alpha\cdot\phi(t,Q^2)}.
\ee 
The shape of $F_{up,lo}(t,Q^2)$ as functions of $t$ depend on the values 
for $t_{i}$ and $f_{i}$. In (\ref{bounds}), $\alpha$ is a scaling constant,
$\beta(t)$ gives roughly the shape of the form factor while the squareroot in
(\ref{bounds}) distinguishes the upper from the lower bound. In the squareroot,
$c(Q^2)$ is a constant in $t$ which depends only on the energy scale $Q^2$,
and $\Delta(t)$
is a function with zeros in every $t_{i}$. The reason why the function
$\Delta(t)$ should behave like this is quite clear: 
since we fixed the value of the form factor at certain points $t_i$, 
the upper and lower bound should coincide in these points
(for an exact definition of the functions appearing in (\ref{bounds}) 
see \cite{hphx9509358}). 

\section{Improving the Bounds}
\label{secimprov}
The method of including `fixed points' (i.e. to fix the value of the
form factor at certain kinematical points) which we summarized in 
\ref{secbounds} is well known and used throughout
the literature. However, this formalism can be applied in a slightly different
way: to include the slope or even higher derivatives of the form factor. 
This is desirable because it is possible to obtain the slope 
e.g.\ for low $t$ from QCD sum rules or for $t$ near the kinematical end point
from the chiral limit. 

Consider two fixed points $t_1$ and $t_2=t_1 + \epsilon$ with $\epsilon$
arbitrary but small. We get the coresponding values of $f(t_i)$ via a
Taylor expansion: 
\be
f(t_1) &\equiv& f_1 \\
f(t_2) &=& f(t_1) + 
           \epsilon \left. \frac{d}{dt}f(t) \right|_{t=t_1} \, . 
\ee
We can now calculate 
the bounds depending on these values for $t_1,t_2,f_1$ and $f_2$ 
and take the limit $\epsilon \to 0$. It is useful to define a function 
$\psi (z,Q^2) \equiv f(z)\phi(z,Q^2)$, in terms of which we obtain 
\be
\label{alptraum}
\psi_{up,lo} &=&
  	\frac{(1-z_1^2)(\psi_1(1-2zz_1+z_1^2)+\psi_1'(z-z_1)(1-z_1^2))}
	{(1 - z z_1)^2} \nonumber\\
&\pm& \frac{(1-z_1^2)(z-z_1)^2}{(1-zz_1)^2\sqrt{1-z^2}} \times \\
&&
	\sqrt{J-\psi_1^2(1+z_1^2)+2\psi_1\psi_1'z_1(1-z_1^2)
	-[\psi_1']^2(1-z_1^2)^2} \nonumber
\ee
where $z_1\equiv z(t_1)$, $\psi_1 \equiv \psi (z_1,Q^2)$ 
and 
$$ 
\psi_1' = \left. \frac{\partial}{\partial z} \psi (z,Q^2) \right|_{z=z_1} \, .
$$

It is obvious how to extend this to higher derivatives (including e.g.\ 
the curvature of the form factor), but the corresponding equations become
tedious. In parts of the numerical analysis presented below the curvature
has been included.  

It has been observed (see \cite{hphx9702300,hphx9705252} and
\cite{hphx9603414,hphx9712417}) that the inequality 
(\ref{main}) also yields restrictions for the derivatives of the form 
factor in the end point $t_-$. We shall use this input to restrict the 
possible values of the form factor at $t_-$ in terms of the slope and 
the curvature. In order to do this we rewrite  (\ref{main}) into the 
form
\be \label{main1}
\frac 1{2\pi}\int_0^{2\pi}\into{\Theta} |\psi(\ex{i\Theta})|^2 \le J(Q^2).
\ee
In the case of $f^0$ it is now possible to
perform a Taylor expansion around $z_0 = 0$ ($t=t_-$) which is convergent 
in the full unit disc, since $f^0$ does not have poles or cuts in this 
region. This is due to the fact that no physical intermediate states can 
contribute. We therefore get the inequality
\be
\label{ineq1}
J(Q^2) \ge\psi^2(0)+\sum_{n=1}^\infty \left(\frac1{n!}\right)^2\psi^{(n)^2}(0).
\ee
where $\psi^{(n)}(0)$ denotes the $n$th derivative of $\psi$ at the point
$z=0$. Note that the form factor, $z(t)$ and thus also $\psi$ are real
in the region $0 \le t \le t_-$.

The form factor $f^+$ is not analytic inside the unit disc, since there 
is a contribution of the $B^*$ in this channel. Hence one expects that the 
Taylor expansion for this form factor converges only within the circle
$0 \le |z| \le |z(m_{B^*}^2)|$. The integral in (\ref{main1}) runs over 
the unit circle and thus the integration contour lies outside the radius 
of convergence of the Taylor series. In order to take into account the 
$B^*$--pole one has to expand in a Laurent series or to  subtract the pole. 
However, closer inspection reveal that these two methods are equivalent.  
We choose to subtract the pole and thus define a function
\be
\tilde f(z) = f(z) - \frac{\mbox{Res}\{f,z_p\}}{z-z_p}\frac{\phi(z_p)}{\phi(z)}
\ee
and obtain
\be
\label{ftilde1}
\int_0^{2\pi}\frac{\into{\Theta}}{2\pi} |\psi(\ex{i\Theta})|^2 
&=& \int_0^{2\pi}\frac{\into{\Theta}}{2\pi} |\tilde \psi(\ex{i\Theta})
+\frac{\mbox{Res}\{\phi f,z_p\}}{\ex{i\Theta} - \ex{i\Theta_p}}|^2 \\
\label{ftilde2}
&=& \frac{\mbox{Res}^2\{\phi f,z_p\}}{1-z_p^2} +
\int_0^{2\pi}\frac{\into{\Theta}}{2\pi} |\tilde \psi(\ex{i\Theta})|^2 
\ee
where we define $\tilde \psi = \tilde f \phi$. Note that the terms 
linear in $\tilde \psi$ and $1/({\ex{i\Theta} - \ex{i\Theta_p}})$ vanish, 
since only terms without $\Theta$ dependence contribute to the integral.  

Thus we obtain instead of (\ref{ineq1}) 
\be 
\label{ineq2}
J(Q^2)-\frac{\mbox{Res}^2\{\phi f,z_p\}}{1-z_p^2} \ge\tilde\psi^2(0)+
\sum_{n=1}^\infty \left(\frac1{n!}\right)^2\tilde\psi^{(n)^2}(0).
\ee

\begin{figure}[htp]
\epsfxsize=9cm
\leavevmode
\centering
\epsffile[70 130 540 650]{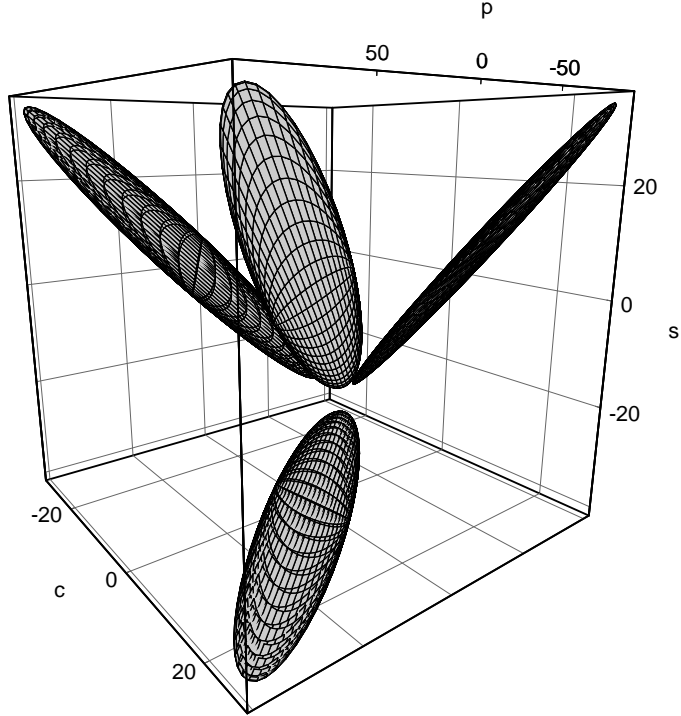} \\
\caption{Possible combinations of point, slope and curvature at the 
end point $f^+(t_-)$, allowed parameter triplets lie inside the ellipsoid. 
To clarify the po\-si\-tion of the ellipsoid, it's contours were drawn on 
three faces of the cube. 
\label{figellipse}}
\end{figure}

If one redefines $J(Q^2)$ to include the residue and if one 
uses the analytic funtion $\tilde f$, (\ref{ineq2}) is the same 
as (\ref{ineq1}).

Since (\ref{ineq1}) is a sum of positive terms, we can 
cut off the sum at some value of $n$ and get (e.g. for $n\le2$)
\be
\label{kurz}
[(\phi f)(0)]^2 + [(\phi f)'(0)]^2 + \frac14 [(\phi f)''(0)]^2 < J(Q^2)
\ee
where we re--substituted $\psi$.

Such equations give us ellipsoids in the parameter space of value, slope
and higher derivatives of the form factor at the kinematical end point, where
all allowed parameter combinations lie inside the ellipsoid. If we take, for
example, $n\le2$ as in (\ref{kurz}) we would get a three--dimensional
ellipsoid in point--slope--curvature--space (see figure \ref{figellipse}).

The resulting constraints are not as good as the ones that can be obtained 
for $B \to D$ transitions \cite{hphx9603414,hphx9712417}; this is due to 
heavy quark symmetries, which are not as useful  in $B \to \pi$ decays.  

However, one may exploit the chiral symmetry for the light degrees of 
freedom as an additional physics input. These symmetries hold at 
small momenta of the pions and are therefore applicable in the end 
point region where the momentum transfer to the leptons becomes maximal. 
One may combine heavy quark and chiral symmetry as in 
\cite{prvxd45x2188}--\nocite{hphx9602353}\cite{hphx9707410} and compute 
the form factors in this limit. One finds \cite{prvxd45x2188}    
\be
\label{chiralform}
f^+(t) &=& \frac{f_B}{2f_\pi}\left[1+\frac{g(M^2-m^2+t)}{M^2+m^2+2\Delta M
-t}\right] \nonumber\\
f^0(t) &=& \frac{f_B}{2f_\pi}\left[1+\frac{g(M^2-m^2+t)}{M^2+m^2+2\Delta M-t}
\right. \\
&&\hspace{1.8cm}
+\left.\frac{t}{M^2-m^2}\left\{1-\frac{g(3M^2-m^2+t)}{M^2+m^2+2\Delta M-t}
\right\}\right] \nonumber
\ee
which results in the residue
\be
\label{chiralres}
\mbox{Res}\{f,t_p\} = -\frac {f_B}{f_\pi}gM(M+\Delta)
\ee
at
\be
t_p = M^2+m^2+2\Delta M
\ee
where $\Delta = m_{B^*} - m_B$. 
The additional parameters which appear in these relations are
the chiral coupling constant $g$ which 
describes the $BB^*\pi$--coupling and the ratio of the decay constants 
$f_B / f_\pi$. 

The form factors (\ref{chiralform}) are valid in the chiral limit, 
which holds only in a small piece of the kinmatically allowed region. 
In order to extend this to the full range, a variety of 
ans\"atze have been invented to extend this pole--behaviour to small
$t$ (see e.g. \cite{zpxc29x637}--\nocite{hphx9401303} \nocite{zpxc48x663}
\cite{prvlx68x2887}); however, this makes the extraction ov $V_{ub}$ 
model dependent. 

We will now use the improved unitarity bounds and 
the chiral limit to derive bounds on the form factors. These bounds 
have the advantage to be model independent, however, as we shall see, 
they are not very tight over the whole range of $t$. This is mainly
due to the fact that the parameters $g$ and $f_B/f_\pi$ suffer from large 
theoretical uncertainties. 

\begin{figure}[htp]
\epsfxsize=9cm
\leavevmode
\centering
\epsffile[70 130 540 650]{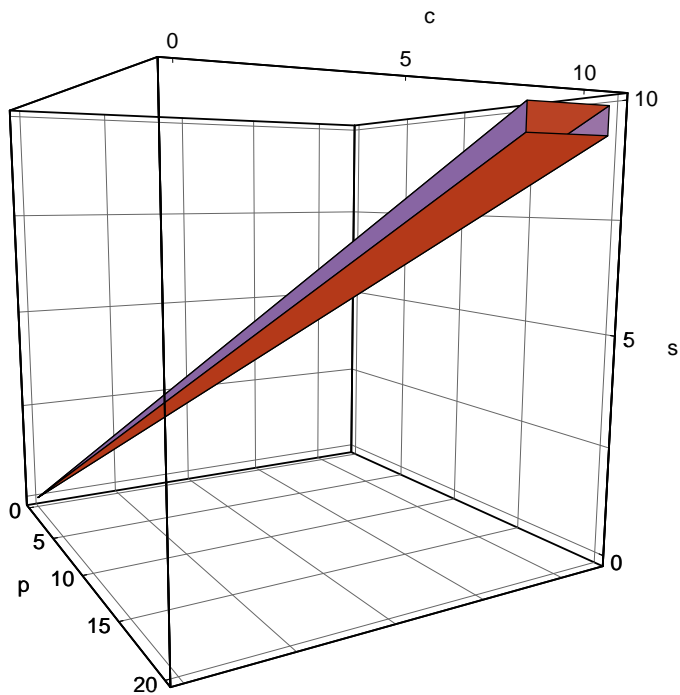} \\
\caption{Possible combinations of point, slope and curvature at the end point
$f^+(t_-)$ using the chiral limit.
\label{figchiralcone}}
\end{figure}

In the following we shall consider $f^+$ only; here we can avoid 
the problem of the large uncertainties in $g$ and $f_B/f_\pi$
by considering fractions like
$f^{+\prime}(t_-)/f^{+} (t_-)$ and $f^{+\prime\prime}(t_-)/f^{+} (t_-)$  
in which the $f_B/f_\pi$--dependence drops out and 
the remaining $g$--dependence is small, since terms involving $g$ 
are suppressed:
\be
\frac{{f^+}'(t_-)}{f^+(t_-)} &=& \frac {M+\Delta}{2 M (M-m) (\Delta+m)}
\frac{1}{1+\frac{\Delta+m}{g(M-m)}} \, , \\
\frac{{f^+}''(t_-)}{f^+(t_-)} &=& \frac {1}{2 M^2 (M-m) (\Delta+m)}
\frac{1}{1+\frac{\Delta+m}{g(M-m)}} \, .
\ee

Thus in the chiral limit the ratios 
$f^{+\prime}(t_-)/f^{+} (t_-)$ and $f^{+\prime\prime}(t_-)/f^{+} (t_-)$  
are practically fixed and we can 
express the slope $s$ and the curvature $c$ in terms of the value $p$ 
of $f^+$ at $t_-$. This would yield a straight line in a $p,s,c$ plot; 
however, varying the coupling $g$ in a generous range between $0.25$ and 
$0.5$ \cite{hphx9605342} and allowing for derivations of the chiral limit
of about 15\%, we obtain the rectangular cone shown in 
figure~\ref{figchiralcone}.       

In case of the residue it is not possible to get rid of the $g$-- and
$f_B/f_\pi$--parameter, so one has to take the full 
uncertainties into account
when calculating the bounds. One could also choose to drop the residue 
completely and work with the redefined $\phi_p(z,Q^2)$ of (\ref{phitrafo}) 
but we decided to use (\ref{chiralres}) because  
the resulting bounds are more stringent.

Using (\ref{chiralform}) and their first derivatives at the end point as
input parameters in (\ref{alptraum}) (the extension to higher derivatives
is trivial) we get upper and lower bounds for $0\le t\le t_-$ which depend
strongly on the chiral parameters $g$ and $f_B/f_\pi$. In addition to
that, a combination of figure \ref{figellipse} and \ref{figchiralcone} can be
used to constrain the allowed range for the end point value of $f(t_-)$ and
its derivatives.

The combination of unitarity bounds, extended to include derivatives, and 
the chiral limit now allows us to calculate bounds for the form factors 
$f^+(t)$ and $f^0(t)$.

\section{Numerical Results and Discussion}
\label{seccalc}
Inserting $t_1=t_-$ ($z_1=0$) in (\ref{alptraum}) one gets
\be
\psi_{up,lo} &=&
  	(\psi_1+\psi_1'z) \pm \frac{z^2}{\sqrt{1-z^2}}
	\sqrt{J-\psi_1^2-[\psi_1']^2}
\ee
which leads to the bounds
\be
\label{endbounds}
F_{up,lo} = \frac{f_1 \phi_1 + (f_1 \phi_1)'z}{\phi(z,Q^2)} \pm
		\frac{z^2}{\phi(z,Q^2)}\sqrt{\frac{J-(f_1 \phi_1)^2 - 
		[(f_1 \phi_1)']^2}{1-z^2}}.
\ee
We used $Q^2=0$ for the QCD calculations and
inserted all possible combinations of end point value and slope into
(\ref{endbounds}). We also allowed 
variation of $g$ and $f_B/f_\pi$ from $0.25\le g\le 0.5$ and 
$1\le f_B/f_\pi\le 1.7$ (\cite{hphx9605342, hlax9710057}). 
Within this variation we determined the bounds 
for fixed parameter values and finally took the loosest ones as our result.

We scanned the parameter space of possible $ps$--pairs by dividing
the allowed $p$--range into small, equally spaced intervals.
For each of these possible $p$--values, there exists an allowed $s$--range
which is also divided into small intervalls. A similar procedure was applied 
for the chiral input parameters.

\begin{figure}[htp]
\epsfxsize=12cm
\leavevmode
\centering
\epsffile{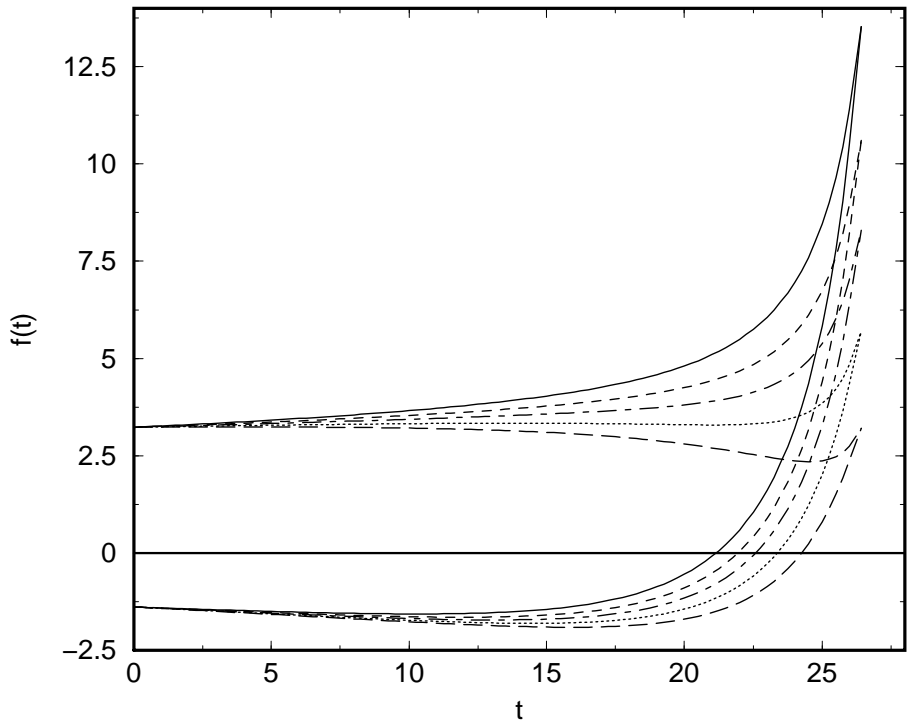} \\
\caption{Bounds on the form factor $f^+(t)$, derived by using a fixed
end point and varying the slope according to the chiral limit
\label{figpieces}} 
\end{figure}

Figure~\ref{figpieces} shows the resulting bounds for the form factor $f^+$. 
The values of the form factor at $t_-$ vary between 3.2 and 13.5 due to the 
uncertainties in $f_B/f_\pi$ and $g$. The two solid lines represent the 
result for the maximal value of the form factor at $t_-$, where the upper 
(lower) line corresponds to the minimal (maximal) possible value for the 
slope at $t_-$. Similarly, we have plotted the corresponding lines for 
intermediate values of the form factor at $t_-$ down to the minimal value, 
represented by the long--dashed line. 

The upper and lower bounds coincide at $t=0$ due to the kinematical
constraint (\ref{kinconst}). We have computed $f^0$ in the same way
as we did for $f^+$, scanning the parameter space for $p$ and $s$ at $t=t_-$. 
The resulting bounds for 
$f^0$ at $t=0$ are much tighter that those for $f^+ (0)$ and hence the 
kinematical constraint has a significant effect on $f^+$.  
In order to be conservative we have varied the parameters for both form
factors independently; that is 
we did not use the correlation between the form factors implied by the 
chiral limit, namely that at $t=t_-$ they are both given by the same two 
parameters $f_B/f_\pi$ and $g$.  

\begin{figure}[htp]
\epsfxsize=12cm
\leavevmode
\centering
\epsffile[70 250 540 540]{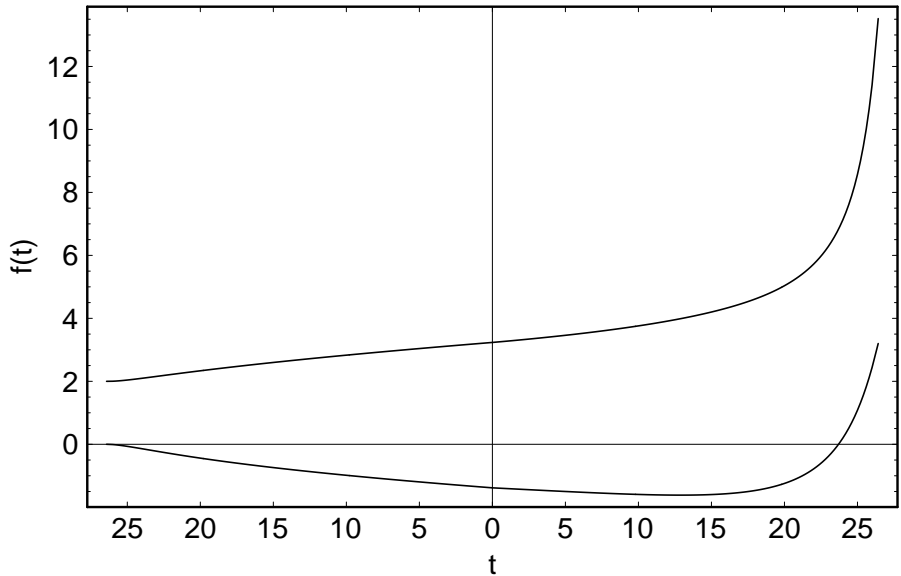} \\
\caption{Bounds on the form factors $f^0(t)$ (on the left side, $t$ increasing 
leftward) and $f^+(t)$ (on the right side, $t$ increasing rightward) derived 
with restriction on possible $ps$--pairs at $f(t_-)$ and including the
kinematical constraint, plotted over the whole
kinematical range of $[0,t_-]$ ($t$ in $\mbox{GeV}^2$). 
\label{figwith}} 
\end{figure}

In figure \ref{figwith} we have combined the results for $f^+$ and $f^0$, 
plotting the upper bound for the maximal values of $f^+ (t_-)$ and 
$f^0 (t_-)$ with the minmal slopes and the lower bounds for the minmal 
values of $f^+ (t_-)$ and $f^0 (t_-)$ with the maximal slopes.  
Compared to the standard method (see figure \ref{figwithout}) 
the inclusion of the slope and the chiral limit has significantly 
improved the bounds. We have to point out, however, that not any 
curve within these two bounds is allowed as a form factor, since QCD 
implies relations between slopes and the form factor values. This can 
be seen in figure~\ref{figpieces}.

Finally one can also use model input into the machinery of unitarity 
bounds. In particular, a model can be used to obtain points and curvatures
away from $t_-$.   

\begin{figure}[hbp]
\epsfxsize=12cm
\leavevmode
\centering
\epsffile[70 250 540 540]{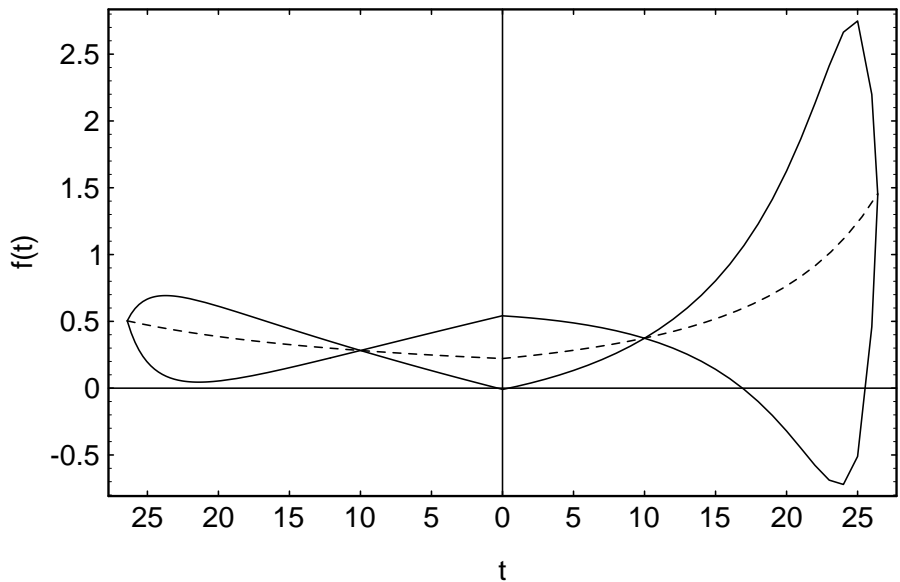} \\
\caption{Bounds on $f^0$ (left) and $f^+$ (right) derived by using the ISGW II
model and using the values at $t=10\mbox{ GeV}^2$ and $t=t_-$ as input
parameters. The bounds are drawn solid, the dashed line corresponds to the
model. 
\label{figmodel2p}} 
\end{figure}

\begin{figure}
\epsfxsize=12cm
\leavevmode
\centering
\epsffile[70 250 540 540]{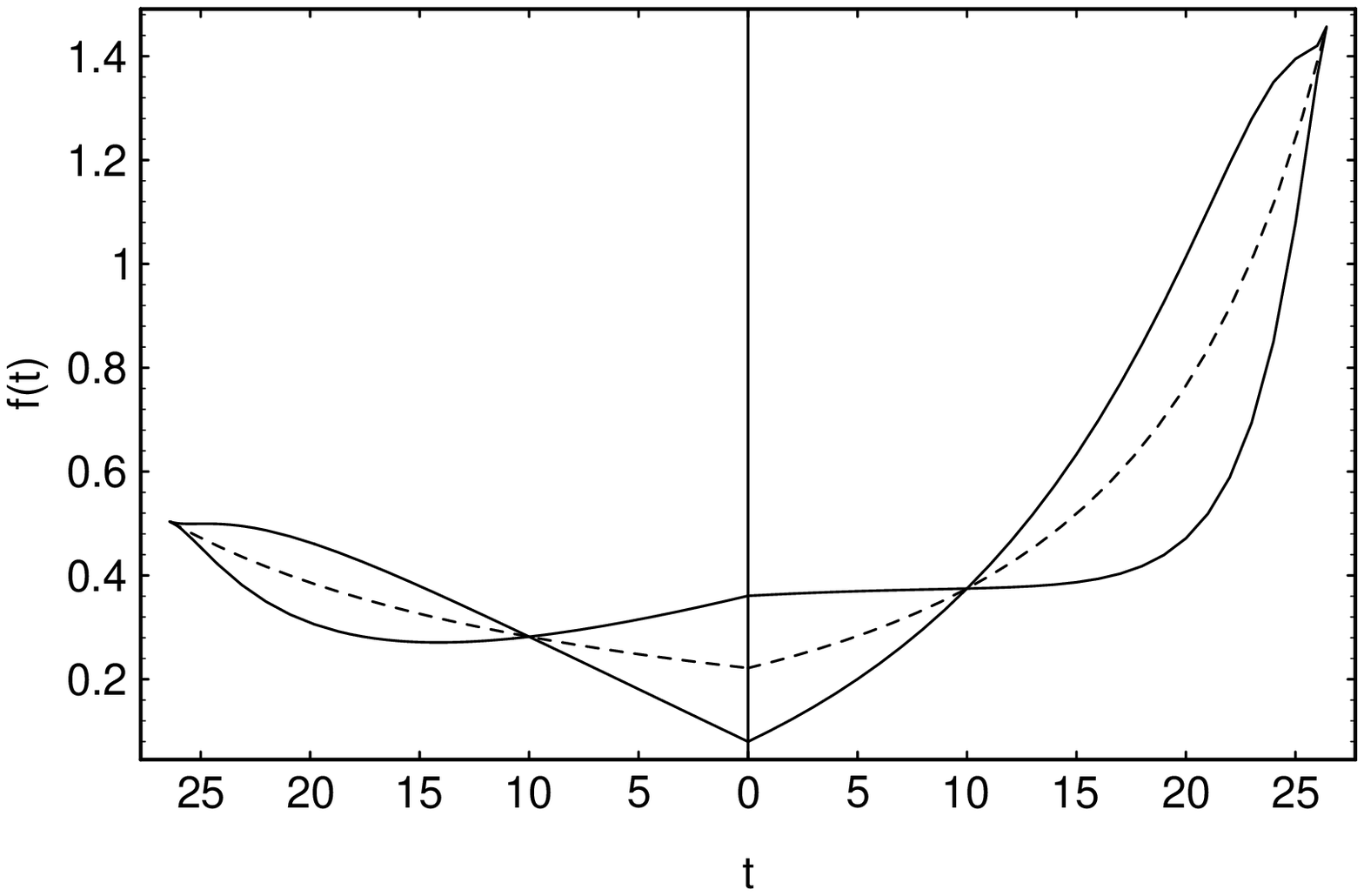} \\
\caption{Same as in figure \ref{figmodel2p} but including the slope at $t=t_-$
as well.
\label{figmodel2a}} 
\end{figure}

In figure \ref{figmodel2p} and \ref{figmodel2a} we have used the 
ISGW II model \cite{hphx9503486} as an input. Fig.\ref{figmodel2p}
shows the bounds obtained by the standard formalism using two 
points ($t=10$ GeV${}^2$ and $t_-$) from this model, while in 
fig.\ref{figmodel2a} we use these two points but also the derivative
at $t_-$, obtained again from ISGW II, as input parameters. The dashed line 
is the model prediction itself. 
 
\begin{figure}[htp]
\epsfxsize=12cm
\leavevmode
\centering
\epsffile[70 250 540 540]{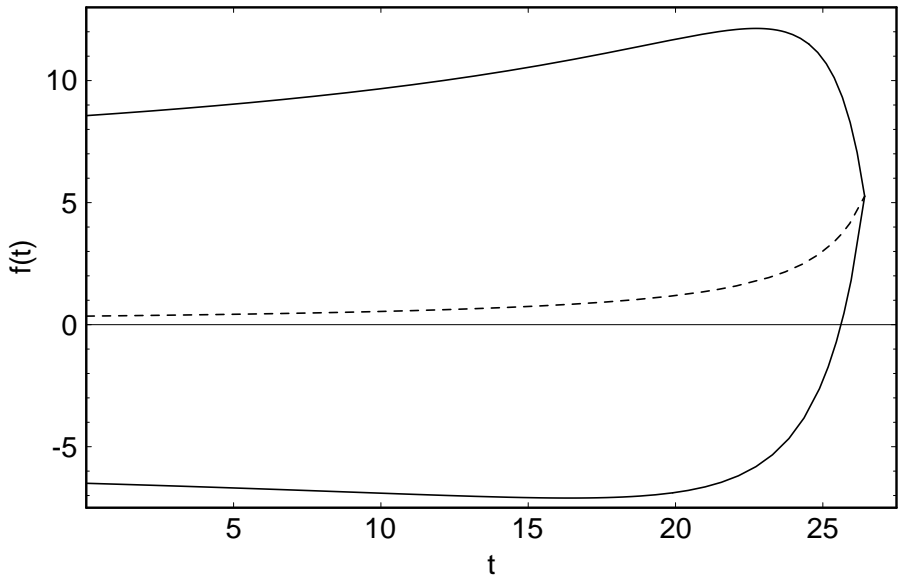} \\
\caption{Bounds on $f^+$ derived by using the BSW model and using the value 
at $t=t_-$ as input parameter. The bounds are drawn solid, the dashed line 
corresponds to the model. 
\label{figpoint}} 
\end{figure}

\begin{figure}[htp]
\epsfxsize=12cm
\leavevmode
\centering
\epsffile[70 250 540 540]{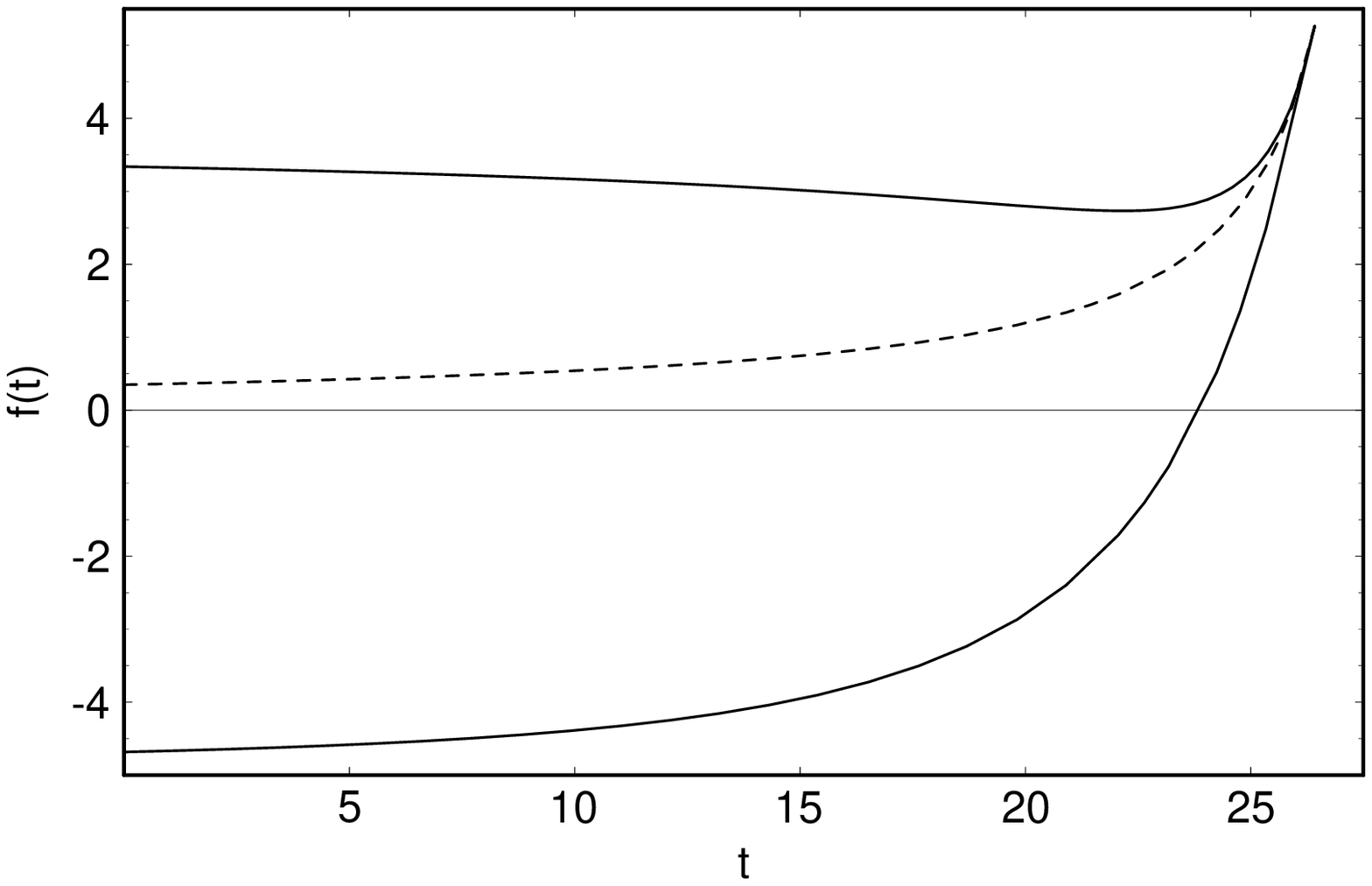} \\
\caption{Same as in figure \ref{figpoint} but including the slope at $t=t_-$
as well.
\label{figslope}} 
\end{figure}

\begin{figure}[htp]
\epsfxsize=12cm
\leavevmode
\centering
\epsffile[70 250 540 540]{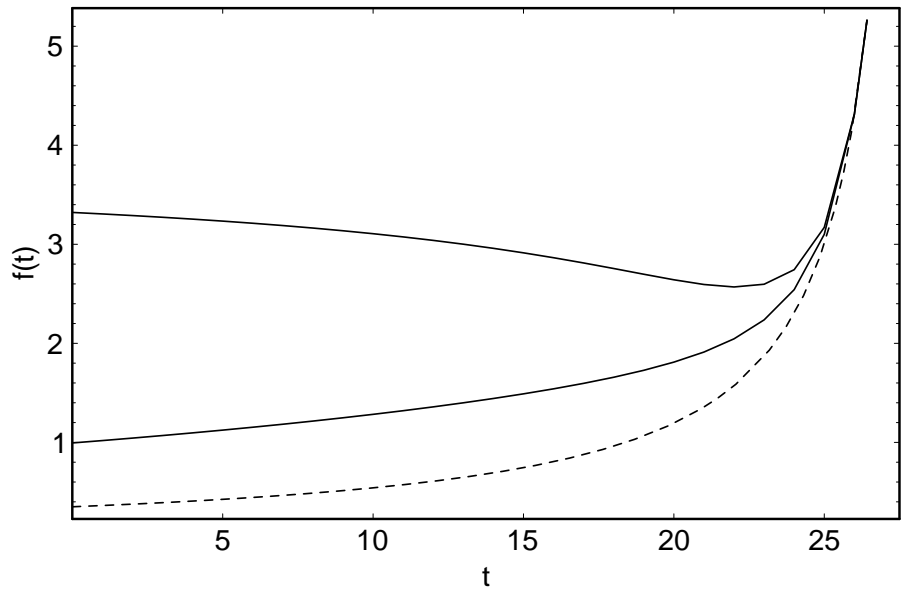} \\
\caption{Same as in figure \ref{figslope} but including the curvature at 
$t=t_-$ as well.
\label{figcurve}} 
\end{figure}

Another commonly used model is the BSW model \cite{zpxc29x637}. 
In fig.\ref{figpoint} we use this model to determine the value of 
the form factor $f^+$ at $t_-$, while in fig.\ref{figslope} also 
the slope at $t_-$ was taken from the model. It is interesting to 
note that if one also includes the curvature of $f^+$ at $t_-$ 
in the BSW model --- as in fig.\ref{figcurve} --- 
the lower bound shifts significantly upwards, such 
that the model becomes inconsistent with the bounds. This indicates that 
the curvature of the BSW model is too large in the end point, since 
the lower bound comes out to be quite high; its value at $t=0$
is about one, which is in contradiction not only with the BSW model, 
but also with sum rule calculations \cite{hphx9305348,hphx9802394}.  
 
The pole form of the BSW model is motivated by the form factor in the 
chiral limit and thus it is no surprise that they have similar curvatures 
of $f^+$ at $t_-$. We take this as an indication 
that one should not exploit the chiral limit to obtain derivatives
of the form factors beyond the slope; going beyond the slope requires
to take into account higher orders in chiral perturabtion theory. 

\section{Conclusions} 
\label{secconc}
Using unitarity and analyticity to derive bounds on form factors 
using input from perturbative QCD has become a widely used tool, 
since this method allows to constrain form factors over the whole
range of $q^2$ in a model independent way. 

This method has attracted some attention in $B$ decays and has been 
combined with heavy quark symmetry; for the $b \to c$ transitions 
this has lead to stringent constraints on the form factors, i.e.
the Isgur--Wise function. 

In $B \to \pi$ transitions heavy quark symmetries are not as useful 
and it is necessary to obtain model independent information from 
other sources. The general bounds in $B \to \pi$ decays 
are not very tight and hence also not very useful. 
Including points where the form factor is known from other sources
improves the bounds; this method has been combined with lattice data
to obtain model independent bounds for the form factors for 
$B \to \pi$ transitions. 

In the present paper we have focussed also at the $B \to \pi$ decays
and improved the bounds on the relevant form factors one more step by 
including slopes and higher derivatives of form factors, which in 
some cases may be known from other sources. Again from unitarity 
the value, the slope, and even higher derivatives of the form factor at 
$t_-$ are constrained to lie inside an ellipsoid which can be used 
as an input into the machinery we have proposed here. 

Still this does not tighten the bounds very much and some more physics
input beyond perturbative QCD is needed. Close to the kinematical end point 
$t_-$ chiral 
perturbation theory is valid and we used the chiral limit of the 
form factors as an input. With this we arrived at bounds which are
much tighter than the general bounds obtained without any knowledge 
on the form factors. The model independent results of our paper are 
shown in fig.\ref{figpieces} and fig.\ref{figwith}.   
 
Finally one may also combine models with the unitarity bounds from QCD. 
Taking points, slopes and curvatures from these models one may test 
the consistency of a given model with QCD. An extensive analysis of 
the various models is beyond the scope of the present paper; we   
only considered the ISGW II and the BSW model as examples. Although 
both models lie within the bounds given in fig.\ref{figwith} (the 
ISGW II touching the lower bound at $t_-$), this still does not 
mean that they are consisten with QCD; the curvature of the BSW model 
at $t_-$ turns out to be incompatibel with the QCD constraints. 

\section*{Acknowledgements}
We thank Changhao Jin, who visited Karlsruhe in the early stage of this
work, for useful discussions on this subject. This work was supported by 
the ``Graduiertenkolleg: Elementarteilchenphysik an Be\-schleu\-ni\-gern''
and the ``Forschergruppe: Quantenfeldtheorie, Computeralgebra und 
Monte Carlo Simulationen'' of the Deut\-sche For\-schungs\-ge\-mein\-schaft.



\begin{thebibliography}{10}

\bibitem{hphx9711268}
C.~G. Boyd and I.~Z. Rothstein,
\newblock Phys. Lett. {\bf B420}, 350 (1998), hep-ph/9711268.

\bibitem{hphx9712399}
S.~Weinzierl and O.~Yakovlev,
\newblock (1997), hep-ph/9712399.

\bibitem{hphx9801421}
D.~S. Hwang and B.-H. Lee,
\newblock (1998), hep-ph/9801421.

\bibitem{hphx9801443}
A.~Khodjamirian and R.~Ruckl,
\newblock (1998), hep-ph/9801443.

\bibitem{prvxd4x3519}
I.~F. Shih and S.~Okubo,
\newblock Phys.~Rev. {\bf D4}, 3519 (1971).

\bibitem{prvxd3x2807}
S.~Okubo,
\newblock Phys.~Rev. {\bf D3}, 2807 (1971).

\bibitem{prvxd4x725}
S.~Okubo,
\newblock Phys.~Rev. {\bf D4}, 725 (1971).

\bibitem{prvxd4x2020}
S.~Okubo and I.~F. Shih,
\newblock Phys.~Rev. {\bf D4}, 2020 (1971).

\bibitem{npxb189x157}
C.~Bourrely, B.~Machet, and E.~de~Rafael,
\newblock Nucl. Phys. {\bf B189}, 157 (1981).

\bibitem{hphx9509358}
L.~Lellouch,
\newblock Nucl. Phys. {\bf B479}, 353 (1996), hep-ph/9509358.

\bibitem{hphx9702300}
C.~G. Boyd and M.~J. Savage,
\newblock Phys. Rev. {\bf D56}, 303 (1997), hep-ph/9702300.

\bibitem{hphx9705252}
C.~G. Boyd, B.~Grinstein, and R.~F. Lebed,
\newblock Phys. Rev. {\bf D56}, 6895 (1997), hep-ph/9705252.

\bibitem{plxb301x257}
J.~G. Korner, D.~Pirjol, and C.~Dominguez,
\newblock Phys. Lett. {\bf B301}, 257 (1993).

\bibitem{hphx9603414}
I.~Caprini and M.~Neubert,
\newblock Phys. Lett. {\bf B380}, 376 (1996), hep-ph/9603414.

\bibitem{hphx9712417}
I.~Caprini, L.~Lellouch, and M.~Neubert,
\newblock (1997), hep-ph/9712417.

\bibitem{prvxd45x2188}
M.~B. Wise,
\newblock Phys. Rev. {\bf D45}, 2188 (1992).

\bibitem{hphx9602353}
G.~Burdman and J.~Kambor,
\newblock Phys. Rev. {\bf D55}, 2817 (1997), hep-ph/9602353.

\bibitem{hphx9707410}
G.~Burdman,
\newblock (1997), hep-ph/9707410.

\bibitem{zpxc29x637}
M.~Wirbel, B.~Stech, and M.~Bauer,
\newblock Z. Phys. {\bf C29}, 637 (1985).

\bibitem{hphx9401303}
B.~Grinstein and P.~F. Mende,
\newblock Nucl. Phys. {\bf B425}, 451 (1994), hep-ph/9401303.

\bibitem{zpxc48x663}
J.~G. Korner, K.~Schilcher, M.~Wirbel, and Y.~L. Wu,
\newblock Z. Phys. {\bf C48}, 663 (1990).

\bibitem{prvlx68x2887}
G.~Burdman and J.~F. Donoghue,
\newblock Phys. Rev. Lett. {\bf 68}, 2887 (1992).

\bibitem{hphx9605342}
R.~Casalbuoni {\em et~al.},
\newblock Phys. Rept. {\bf 281}, 145 (1997), hep-ph/9605342.

\bibitem{hlax9710057}
J.~M. Flynn and C.~T. Sachrajda,
\newblock (1997), hep-lat/9710057.

\bibitem{hphx9503486}
D.~Scora and N.~Isgur,
\newblock Phys. Rev. {\bf D52}, 2783 (1995), hep-ph/9503486.

\bibitem{hphx9305348}
V.~M. Belyaev, A.~Khodjamirian, and R.~Ruckl,
\newblock Z. Phys. {\bf C60}, 349 (1993), hep-ph/9305348.

\bibitem{hphx9802394}
P.~Ball,
\newblock (1998), hep-ph/9802394.

\end{thebibliography}


\mbox{}
\newpage

\end{document}